\documentclass[aps,showpacs,preprintnumbers,amsmath,twocolumn, tightenlines]{revtex4}

\usepackage{bm}
\usepackage{longtable}
\usepackage{amsmath}
\usepackage{dcolumn}
\usepackage{placeins}

\begin{document}

\bibliographystyle{apsrev}

\title {On the prospects of building optical atomic clocks using Er I or Er III.}

\author{A. Kozlov}\email[email:]{o.kozloff@student.unsw.edu.au}
\author{V. A. Dzuba}\email[email:]{dzuba@phys.unsw.edu.au}
\author{V. V. Flambaum}\email[email:]{flambaum@phys.unsw.edu.au}

\affiliation{School of Physics, University of New South Wales, Sydney
  2052, Australia}

    \date{\today}

    \begin{abstract}
The possibility of using neutral and double ionized erbium for atomic clocks of high precision is investigated. In both cases the narrow electric quadrupole clock transition between the ground and first exited state of the same configuration lies in optical region. The estimated ratio of decay width to transition energy is less then $10^{-20}$. We demonstrate that this transitions are not sensitive to black body radiation and if other perturbations are also considered the relative accuracy of the clocks can probably reach the level of $10^{-18}$ or better.
    \end{abstract}


\maketitle

\section{Introduction}
Atomic clocks are widely used in many areas of science and industry due to their extremely high accuracy. Cesium primary frequency standard which
is currently used to define the SI units of time and length  has
fractional accuracy  of  $10^{-16}$~\cite{NIST}. Frequency standards based on optical transitions in neutral atoms trapped in optical lattice aim at fractional accuracy of $10^{-18}$~\cite{Katori}. The use of
nuclear optical transition or optical transitions in highly charged ions may bring the accuracy down to the level of $10^{-19}$~\cite{Th,HCI1,hole,Cf,IonClock,NdSm}. Such 
improvement in accuracy is important since it widens the area of scientific research and technical applications. For example, the fractional accuracy of $10^{-18}$ to $10^{-19}$ is needed to test the so-called $\alpha$-dipole hypothesis which claims that the fine structure constant $\alpha$ is different in different places of Universe, changing smoothly along a certain direction in space~\cite{dipole}.

It was suggested in Ref.~\cite{IonClock,f12} to use highly charged ions with the $4f^{12}$ configuration of valence electrons (isoelectronic sequence from Os$^{18+}$ to U$^{34+}$). Electric quadrupole transition between the ground and first excited states of this configuration is very narrow and not sensitive to external perturbations. It was shown in Ref~\cite{IonClock,f12} that these transitions in highly charged ions can be used for atomic optical clocks of extremely high fractional accuracy of about  $10^{-18}$. 

In this paper we study similar transitions in neural erbium and its doubly ionized ion. The electron configuration is  $\left[\text{Xe}\right]4f^{12}6s^2$ for Er~I and $\left[\text{Xe}\right]4f^{12}$ for Er~III. Both systems have the $^3$H$_6$ ground state and $^3$F$_4$ first excited state, the same as in ions considered in Ref.~\cite{IonClock,f12}. It is natural to expect that many features of these transitions would be very similar to those of the highly charged ions. The main difference is expected to come from the larger polarizabilities. The values of the electric dipole polarizabilities of highly charged ions are small due to their small size and sparse spectrum~\cite{IonClock}. This makes the ion transitions to be insensitive to the frequency shift due to black body radiation (BBR). In neutral atoms polarizabilities are large and BBR shift is often the main factor limiting the accuracy of the microwave or optical clocks. It is important that the BBR shift is proportional to the difference of the polarizabilities of two clock states. When the states are sufficiently different (e.g. belong to different configurations) the difference is of the same order of magnitude as the polarizabilities. This is the case for most of the optical clock transitions considered so far. Most of optical clocks use the $ns^2 \ ^1$S$_0 - nsnp \ ^3$P$_0$ transition in two-valence-electron atoms (Sr, Yb, Hg, etc., see e.g.~\cite{Katori}). In these atoms the polarizability of the upper state is about two times larger than for the lower state~\cite{Mitroy}. In contrast, the transitions considered in present paper are between states of the same configuration. It turns out that due to the similarities in the wave function of both states the polarizabilities are almost the same, the difference is about four orders of magnitude smaller that the polarizabilities. This makes the transitions to be insensitive to the BBR frequency shift. 

The most obvious disadvantage of the use of the $^3$H$_6$ - $^3$F$_4$ transition as a clock transition is high value of the total angular momentum of both states. This makes the transitions to be sensitive to the gradients of electric field. Corresponding uncertainty can be significantly reduced by averaging over transitions with different projections of the total angular momentum. We consider this and other perturbations in this paper and demonstrate that the fractional accuracy of $10^{-18}$ and lower is probably possible for the Er~I and Er~III clock transitions. 

\section{Clock transition}
First excited state for both Er and Er$^{2+}$ is $^3$F$_4$ with total angular momentum $J_e=4$, while the ground state is $^3$H$_6$ with  $J_g=6$. Both ground and first excited states belong to the same electron configuration ($4f^{12}$ in Er~III and $4f^{12}6s^2$ in Er~I). The first non-vanishing amplitude of transition between them is electric quadrupole transition (E2). The decay width of E2 transition is given by the following expression

\begin{equation}\label{E2Width}
\Gamma_e=\frac{1}{15}\alpha^5\omega^{5}\frac{\langle e\left|\left|E2\right|\right|g\rangle^2}{2J_e+1}, 
\end{equation}
where $\omega$ is the transition frequency, $\alpha=1/137$ is the fine structure constant, $J_e$ is the total moment of exited state, $E2$ is the operator of electric quadrupole transition ($r^2Y_{2m}(\theta,\phi)$). Here and further in the paper we employ atomic units as a default system of units.

Table \ref{Tab:1} represents some important properties of neutral and doubly-ionized erbium atoms. Energies are taken from the NIST database~\cite{NISTe}, lifetimes are calculated using configuration interaction (CI) method described in \cite{Dzuba:2008, Flambaum:2008}. Comparing these properties with the ones of highly charged ions \cite{f12} one can notice that the quality factors $Q$ ($Q=\omega/\Gamma$) for Er~I and Er~III are of the same order as for the highly charged ions. This is because the frequency of the clock transition in Er~I and Er~III is about two times smaller than in the ions~\cite{f12}. Since frequency enters the transition probability (\ref{E2Width}) in power five, the resulting factor of about 30 compensates for the smaller E2 transition amplitudes in ions.

\begin{table}[tb]\center
\caption{Characteristics of proposed clock transitions in neutral and double ionized Er. Numbers in square brackets represent the power of 10.}
{\renewcommand{\arraystretch}{0}%
\begin{tabular}{l c c c c c}
\hline\hline
\rule{0pt}{4pt}\\
Atom & $\Delta E,$ & $\lambda,$ & $\Gamma,$ & $\tau,$ & $1/Q$ \\
& $\text{cm}^{-1}$ & $\text{nm}$ & $\mu\text{Hz}$ & $\text{hours}$\\
\rule{0pt}{4pt}\\
\hline
\rule{0pt}{4pt}\\
Er & 5035 & 1986 & 47 & 5.9 & 4.7[-20]\\
\rule{0pt}{2pt}\\
Er$^{2+}$ & 5081 & 1966 & 11 & 24 & 1.2[-20]\\
\rule{0pt}{4pt}\\
\hline\hline
\end{tabular}}\label{Tab:1}
\end{table}
\FloatBarrier

Systematic effects which limit the accuracy of atomic clock include BBR, interaction of atomic quadrupole moments with gradients of electric field, micro and secular motion, Stark and Zeeman shifts, background-gas collisions, gravitational shift, etc. Some of these factors were discussed in \cite{Chou:2010, Derevianko:2012}. The most significant factors are BBR, quadrupole and  Zeeman shifts.

The BBR shift originates from perturbation of the clock transition by the environment photon bath. This shift is described by the following equation \cite{Porsev:2006}
\begin{equation}\label{BBR}
\frac{\Delta\omega}{\omega_0}\left|_{\text{\parbox{.3in}{BBR}}}\approx -\frac{2\pi^3\alpha^3}{15}\frac{T^4}{\omega_0}\Delta\alpha(0)\right. \equiv \beta\left(\frac{T}{300K}\right)^4,
\end{equation}
where $T$ is the temperature. The differential scalar polarizability can be calculated using equation 
 \begin{align}\label{eq:da}
\Delta\alpha(0)&=\frac{2}{3(2J_e+1)}\sum_k\frac{\langle k||\hat{\textbf{d}}||  e \rangle^2}{E_e-E_k}\\
&-\frac{2}{3(2J_g+1)}\sum_k\frac{\langle k||\hat{\textbf{d}}|| g \rangle^2}{E_g-E_k}\nonumber
\end{align}
where $\hat{\textbf{d}}=-e\hat{\textbf{r}}$ is the dipole moment operator.
This difference must be very small for the states considered in present work. The states belong to the same configuration ($4f^{12}6s^2$ in Er~I and $4f^{12}$ in Er~III) and if we neglect small differences in the relativistic composition of the states (relative contributions of the $4f_{5/2}$ and $4f_{7/2}$ states) as well as small differences in mixing with other configurations, then the clock states differ by angular part of the wave functions which cannot affect the values of scalars like polarizability, energy, etc. Indeed, the CI calculations show that the difference in energies of the clock states (5035 cm$^{-1}$ in Er~I and 5082 cm$^{-1}$ in Er~III) is only about $10^{-4}$ of the total energy of the corresponding configuration. The difference in polarizabilities is also expected to be small. It is instructive to consider a particular example in detail.

Let's consider clock transitions of Er~I in single configuration approximation. The wave function of the ground state can be written as
\begin{equation}
\Psi_g = |4f^{12} \ ^3{\rm H}_6\rangle | 6s^2 \ ^1{\rm S}_0 \rangle.
\label{eq:g}
\end{equation}
The wave function of the upper clock state is
\begin{equation}
\Psi_e = |4f^{12} \ ^3{\rm F}_4\rangle | 6s^2 \ ^1{\rm S}_0 \rangle.
\label{eq:e}
\end{equation}
The expressions for polarizabilities (\ref{eq:da}) are strongly dominated by transitions to specific states of the $4f^{12}6s6p$ configuration. The states which contribute the most to the polarizability of the ground state (\ref{eq:g}) are
\begin{equation}
\Psi_{JM} = C^{JM}_{6M10}|4f^{12} \ ^3{\rm H}_6\rangle | 6s6p \ ^1{\rm P}_1 \rangle, \ J=5,6,7.
\label{eq:Jg}
\end{equation}
Here $J$ is total angular momentum, $M$ is its projection, and $C^{JM}_{6M10}$ is the Clebsch-Gordan coefficient.
Transitions from the clock state (\ref{eq:e}) are dominated by
\begin{equation}
\Psi_{J^{\prime}M^{\prime}} = C^{J^{\prime}M^{\prime}}_{4M^{\prime}10}|4f^{12} \ ^3{\rm F}_4\rangle | 6s6p \ ^1{\rm P}_1 \rangle, \ J^{\prime}=3,4,5.
\label{eq:Je}
\end{equation}
Experimental and calculated energies and g-factors for six states (\ref{eq:Jg},\ref{eq:Je}) as well as calculated electric dipole transition amplitudes from the ground state (\ref{eq:g}) to excited odd states (\ref{eq:Jg}) and from second clock state (\ref{eq:g}) to odd states (\ref{eq:Je}) are presented in Table~\ref{tab:2}. Note that the term notations in the Table are taken from the NIST database. We believe that they are not accurate and that $^3$P should be replaced by $^1$P for all states in the Table. Note also that the energies within each triplet are very close and the difference between lower and upper triplet energies is very close to the energy difference between clock states. This is natural because the groups of states differ only by the configuration of $4f$ electrons ($^3$H$_6$ or $^3$F$_4$). Finally note that in the considered approximation the sum over $k$ in (\ref{eq:da}) is reduced to the sum over $J$ (as in (\ref{eq:Jg})) for the first term and over $J^{\prime}$ (as in (\ref{eq:Je})) for the second term. The above mentioned notion about equal energy shifts means that energy denominators in both terms are almost the same. 

Reduced matrix elements of the electric dipole transitions in (\ref{eq:da}) can be written using (\ref{eq:g},\ref{eq:e},\ref{eq:Jg},\ref{eq:Je}) as
\begin{eqnarray}
\label{rme1}
\langle \Psi_g ||\hat{\textbf{d}}|| \Psi_{J} \rangle &=& (-1)^{6-M} \left( \begin{array}{rrr} 6 & 1 & J \\ -M & 0 & M \\ \end{array} \right)^{-1}\\  &\times&C^{JM}_{6M10} \langle 6s^2 \ ^1{\rm S}_0 |\hat{\textbf{d}}|6s6p ^1{\rm P}_1 \rangle, \nonumber \\
\label{rme2}
\langle \Psi_e ||\hat{\textbf{d}}|| \Psi_{J^{\prime}} \rangle &=& (-1)^{4-M^{\prime}} \left( \begin{array}{rrr} 4 & 1 & J^{\prime} \\ -M^{\prime} & 0 & M^{\prime} \\ \end{array} \right)^{-1}\\  &\times&C^{J^{\prime}M^{\prime}}_{4M^{\prime}10} \langle 6s^2 \ ^1{\rm S}_0 |\hat{\textbf{d}}|6s6p ^1{\rm P}_1 \rangle. \nonumber 
\end{eqnarray}
Note that matrix elements of electric dipole operator are the same in (\ref{rme1}) and (\ref{rme2}). 

Substituting (\ref{rme1}) and (\ref{rme2}) into (\ref{eq:da}) and neglecting small difference in energy denominators we reduce summation over $J$ and $J^{\prime}$ to summation over angular coefficients. This summation gives the same result for both clock states:
\begin{eqnarray}
&&\frac{2}{3(2J_g+1)}\sum_J \left( \begin{array}{rrr} J_g & 1 & J \\ -M & 0 & M \\ \end{array} \right)^{-2} \left(C^{JM}_{J_gM10}\right)^2 = \\
&&\frac{2}{3(2J_e+1)}\sum_{J^{\prime}} \left( \begin{array}{rrr} J_e & 1 & J^{\prime} \\ -M^{\prime} & 0 & M^{\prime} \\ \end{array} \right)^{-2} \left(C^{J^{\prime}M^{\prime}}_{J_eM^{\prime}10}\right)^2 =2, \nonumber \\
&&(J_g = 6, \ J_e = 4). \nonumber
\end{eqnarray} 
Therefore, in this approximation the polarizabilities of both clock states are exactly the same.

\begin{table}[tb]\center
\caption{Strong E1 transitions from clock states in Er. Energies are measured from the ground state. Experimental values for energies and g-factors are taken from the NIST database~\cite{NISTe}. Calculated energies and g-factors are presented in parentheses. Reduced matrix elements of E1 transitions (RME) were calculated using CI method.}
{\renewcommand{\arraystretch}{0}%
\begin{tabular}{c|c|c|c|c}
\hline\hline
\rule{0pt}{2pt} & \rule{0pt}{2pt} & \rule{0pt}{2pt} & \rule{0pt}{2pt} & \rule{0pt}{2pt} \\
Term & J & Energy & g-factor & RME \\
     &   & (cm$^{-1}$) &  & (a.u.)\\
\rule{0pt}{2pt} & \rule{0pt}{2pt} & \rule{0pt}{2pt} & \rule{0pt}{2pt} & \rule{0pt}{2pt} \\
\hline
\multicolumn{5}{c}{\rule{0pt}{4pt}}\\
\multicolumn{5}{c}{Ground state $4f^{12}(^3\text{H})6s^2(^1\text{S}), \ J=6, \ E=0$}\\
\multicolumn{5}{c}{\rule{0pt}{4pt}}\\
\hline
\rule{0pt}{2pt} & \rule{0pt}{2pt} & \rule{0pt}{2pt} & \rule{0pt}{2pt} & \rule{0pt}{2pt} \\
$4f^{12}(^3\text{H})6s6p(^3\text{P})$ & 7 & 25598(25530) & 1.15(1.14) & 14.70\\
$4f^{12}(^3\text{H})6s6p(^3\text{P})$ & 6 & 26237(26217) & 1.16(1.16) & 13.77\\
$4f^{12}(^3\text{H})6s6p(^3\text{P})$ & 5 & 25364(25445) & 1.18(1.19) & 12.55\\
\rule{0pt}{4pt} & \rule{0pt}{4pt} & \rule{0pt}{4pt} & \rule{0pt}{4pt} & \rule{0pt}{4pt}\\
\hline
\multicolumn{5}{c}{\rule{0pt}{4pt}}\\
\multicolumn{5}{c}{Exited state (J=4) $4f^{12}(^3\text{F})6s^2(^1\text{S}), \ J=4, \ E=5035 \ \text{cm}^{-1}$}\\
\multicolumn{5}{c}{\rule{0pt}{4pt}}\\
\hline
\rule{0pt}{2pt} & \rule{0pt}{2pt} & \rule{0pt}{2pt} & \rule{0pt}{2pt} & \rule{0pt}{2pt} \\
$4f^{12}(^3\text{F})6s6p(^3\text{P})$ & 5 & 31364(31903) & 1.23(1.07) & 12.52\\
$4f^{12}(^3\text{F})6s6p(^3\text{P})$ & 4 & 31155(31883) & 1.14(1.11) & 11.12\\
$4f^{12}(^3\text{F})6s6p(^3\text{P})$ & 3 & 31364(31917) & 1.23(1.11) & 10.03\\
\rule{0pt}{4pt} & \rule{0pt}{4pt} & \rule{0pt}{4pt} & \rule{0pt}{4pt} & \rule{0pt}{4pt}\\
\hline\hline
\end{tabular}}\label{tab:2}
\end{table}

For further discussion of scalar polarizabilities calculation of Er and Er$^{2+}$ it is convenient to employ secondary quantization formalism and rewrite the wave functions in terms of hole states. In these notations the wave functions of clock states for Er~I and Er~III are practically the same and indicate strong domination of the $4f^2_{7/2}$ configuration.
\begin{eqnarray}
\label{decompose}
\Psi_{66}&=&0.95|4f_{\frac{7}{2};\frac{7}{2}}4f_{\frac{7}{2};\frac{5}{2}}\rangle +0.31|4f_{\frac{7}{2};\frac{7}{2}}4f_{\frac{5}{2};\frac{5}{2}}\rangle\\ \Psi_{44}&=&0.81|4f_{\frac{7}{2};\frac{5}{2}}^14f_{\frac{7}{2};\frac{3}{2}}^1\rangle +0.55|4f_{\frac{7}{2};\frac{7}{2}}^14f_{\frac{7}{2};\frac{1}{2}}^1\rangle + \dots \nonumber \end{eqnarray}
Here $\Psi_{JM}$ is the wave function of the state with total angular momentum $J$ and its projection $M$. In the expression for $\Psi_{44}$ there are four more terms of the $4f_{7/2}4f_{5/2}$ and $4f^2_{5/2}$ configurations. One can see from these expressions that the $4f^2_{7/2}$ configuration contributes 90\% to the first state and 96\% to the second state. The difference in energy and polarizabilities of these states is due to this small difference in the composition of the wave functions. In addition to this the difference in the $4f_{7/2}$ and $4f_{5/2}$ wave functions is small due to strong suppression of the relativistic effects for states with high angular momentum.

Above we considered clock states in a single configuration approximation. Adding more configurations lead to the following composition of the states:
\begin{align}
&\text{Er:}& (J=6)\,& 4f^{12}6s^{2} - 93.489\%\\
& & & 4f^{12}6p^{2} - 5.763\%\nonumber\\
& & (J=4)\,& 4f^{12}6s^{2} - 93.497\%\nonumber\\
& & & 4f^{12}6p^{2} - 5.763\%\nonumber\\
&\text{Er$^{2+}$:}& (J=6)\,& 4f^{12} - 99.9\%\nonumber\\
& & (J=4)\,& 4f^{12} - 99.9\%\nonumber
\end{align}
The admixture of other configurations is small. We see that the composition of clock states is almost identical for both Er~I and Er~III. Therefore, after adding more configurations the difference in polarizabilities of clock states remains small. The CI calculations in single configuration approximation gives the following values for the polarizabilities: $\alpha_0 \approx 250$ a.u. for Er~I, $\alpha_0 \approx 10$ a.u. for Er~III while $\Delta \alpha_0 \approx 0.05$ a.u. for both Er~I and Er~III. This leads to small value of the BBR shift parameter $\beta$ (see (\ref{BBR})): $\beta \approx 10^{-18}$ for both Er~I and Er~III. Note that our values for the polarizabilities are probably overestimated. The value for Er~I is about 1.6 times larger than the value 153(38) a.u. presented in Ref.~\cite{Miller}. Adding more configurations reduces the value of $\alpha_0$ while having little effect on $\Delta \alpha_0$. Accurate calculations of the polarizabilities for erbium is problematic due to complicated electron structure. However, we believe that the small value for the difference in polarizabilities of the clock states is reliable because of well established similarities between the states.
The value $\beta = 10^{-18}$ means that the BBR shift $\Delta \omega/\omega = 10^{-18}$ at room temperature. Better accuracy might be possible if cooling is used~\cite{Katori1, Chou:2010}.

Building atomic clock with neutral erbium would involve capturing the atoms in optical lattice. Then the frequency of the clock transition would be affected by the lattice electric field. The standard way around this problem is finding the {\em magic} wavelength for the lattice field~\cite{Katori} so that the energy shifts of both states are exactly the same. On our present level of accuracy we cannot reliably calculate accurate values of the magic frequencies. However, it is easy to prove their existence and to indicate their positions approximately. The polarizabilities are smooth and monotonic functions of frequencies apart from small areas near resonances. Since two states are slightly different, they have resonances at different frequencies. As soon as one of the polarizabilities of two states has a resonance while the other one does not, there is a value of frequency for which two polarizabilities have the same value. This happens in the vicinity of every transition from the ground state to odd states with total angular momentum $J=5,6,7$ and every transition from second clock state to odd states with $J=3,4,5$. For example, for states in Table~\ref{tab:2} there are magic frequencies near $E=25364 \ {\rm cm}^{-1}$,                 $E=25598 \ {\rm cm}^{-1}$, $E=26120 \ {\rm cm}^{-1}$, $E=26237 \ {\rm cm}^{-1}$, and $E=26329 \ {\rm cm}^{-1}$.

Coupling of the atomic quadrupole moment to the gradient of external electric field is another important source of systematic frequency shift.
The shift is given by
\begin{equation}\label{qaudr}   
H_Q=\frac{1}{2}Q\frac{\partial E_z}{\partial z},
\end{equation}
where $Q$ is the quadrupole moment of the clock state. The single configuration CI calculations give practically the same values of the quadrupole moments for clock states of Er~I and Er~III: $Q_6 = 0.66$ a.u., $Q_4=-0.57$ a.u. Using $\partial E_z/\partial z \sim 10^{6} \ {\rm V/m}^2$ as a typical value for the traps~\cite{Barwood:2004} leads to $\Delta \omega/\omega \sim 10^{-15}$ for Er~III. Using $\partial E_z/\partial z \sim 10^{7} \ {\rm V/m}^2$ as a typical value for optical lattice \cite{Porsev:2004} leads to $\Delta \omega/\omega \sim 10^{-14}$ for Er~I. These shifts are large and need to be suppressed or canceled out if possible. There are several ways to achieve this \cite{Itano:2000, Dube:2005, Roos:2006}. It was suggested in Ref.~\cite{f12} to recover the actual frequency of the clock from measuring two frequencies of the transitions between states of different projections of the total angular momentum $J$. This method required the knowledge of the ratio of quadrupole moments $Q_6/Q_4$ of the clock states. It can be used for Er~I and Er~III as well. However, there is another possibility which may give more accurate results. Measuring three frequencies instead of two allows to exclude the ratio of quadrupole moments so that the result would not depend on the uncertainty in this ratio.

Energy shift for the state with total angular momentum $J$ and its projection $M$ can be written as
\begin{equation}\label{Qshift}
\delta E_{JM}\sim \frac{3M^2-J(J+1)}{3J^2-J(J+1)}Q_J\frac{\partial E_z}{\partial z}=C_{JM}Q_J\frac{\partial E_z}{\partial z}
\end{equation}

Using (\ref{Qshift}) one can write down the expression for the frequency of transition between two levels $J_1, M_1\rightarrow J_2, M_2$ as

\begin{equation}\label{omega0}
\omega=\omega_0+(C_{J_1,M_1}Q_{J_1}-C_{J_2,M_2}Q_{J_2})\frac{\partial E_z}{\partial z}.
\end{equation} 
Writing similar expressions for transitions between states of different projections $M'_1, m'_2, M''_1, M''_2$ and solving the resulting system of linear equations leads to
\begin{equation}
\omega_0=\begin{vmatrix}
\omega\text{\,} & C_{J_1,M_1} & C_{J_2,M_2}\\
\omega' & C_{J_1,M_1'} & C_{J_2,M_2'}\\
\omega'' & C_{J_1,M_1''} & C_{J_2,M_2''}
\end{vmatrix}/\begin{vmatrix}
1 & C_{J_1,M_1} & C_{J_2,M_2}\\
1 & C_{J_1,M_1'} & C_{J_2,M_2'}\\
1 & C_{J_1,M_1''} & C_{J_2,M_2''}
\end{vmatrix}
\end{equation}

The above expression allows to exclude large quadrupole shift from three known transitions with different magnetic quantum numbers. 

First order Zeeman shift can be canceled out by averaging results for two different frequencies that corresponds to transitions $J_1, M_1\rightarrow J_2, M_2$ and $J_1,-M_1\rightarrow J_2,-M_2$. This is well known technique that was described in details in \cite{Rosenband:2007, Bernard:1998}. It is also possible to perform accurate estimate of second order Zeeman effect (see for example \cite{Chou:2010}). It should be noted that for even isotopes of erbium, Zeeman effect should be at least several orders of magnitude smaller compared to Al$^+$ ion~\cite{Chou:2010} due to absence of hyperfine structure.

Micromotion and secular motion can also cause significant systematic shift of the clock frequency via special relativity effect known as time dilation~\cite{Hume:2010}:
\begin{equation}
\frac{\Delta\omega}{\omega_0}\left|_{\text{Time Dilation}}\sim-\frac{3T}{2Mc^2}\right.,
\end{equation} 
where $T$ is effective temperature of ion motion in atomic units. This shift is suppressed in erbium due to its relatively large mass. Taking kinetic temperature of the cooled ion $T=2$ mK as was achieved for strontium ion clock \cite{Madej:2012}, one obtains $1.8\times10^{-18}$ for time dilation shift. Micromotion also causes Stark shift in ion clocks. It was shown in \cite{Berkeland:1998, Madej:2004, Madej:2012} that this shift can be canceled by the time dilation shift by an appropriate choice of the angular frequency of the trap field. At least one order of magnitude suppression up to $10^{-19}$ is expected to be a result of such cancellation for proposed clock.  

\section{Conclusion}

We show that neutral Er I and double ionized Er III are promising candidates for optical atomic clocks. Both systems are not sensitive to BBR shift due to extremely small differential scalar polarizability. Dominating systematic shift comes from coupling of the atomic quadrupole moments to the gradients of electric field. However, this shift can be strongly suppressed by averaging over transitions with different projections of the total angular momentum. Other systematic shifts are either small or can be suppressed. The fractional accuracy of $10^{-18}$ is probably achievable for both types of clocks.

\acknowledgments

The authors are grateful to M. G. Kozlov and J. Berengut for useful discussions.
The work was supported in part by the Australian Research Council.

\end{document}